# Sociological Explanation of Fear of Crime in Public Spaces Case Study Mashhad

Mahdie Rezvani[1], Yaser Sadra[2]

[1]Department of Sociology, Ferdowsi University of Mashhad (FUM), Mashhad, Iran.
Email: ma_re79@mail.um.ac.ir

[2]Department of computer sciences, Shandiz Institute of Higher Education, Mashhad, Iran.
Email: y.sadra@shandiz.ac.ir

*Abstract*

Fear of crime is an especially a problem that has troubled the urban communities, affected the urban dissatisfaction and many factors in the city have created and intensified it. Fear of crime is of important issues which reduces access to public places and restricts interaction with these places.
 The research method is surveying and the information collection technique is through questionnaire. Probable span sampling (PPS) method is used. Sample population was 2000 households which were selected randomly in five categorical clusters from Mashhad city. In this research we try In addition to investigation social factors, it has been tried in the research that spatial components affecting fear of crime in public places inside and outside the neighborhoods is considered.

## 1. Introduction

Fear of crime is a social reality (Reid and Roberts, 1998:313) and it is specially a problem that has troubled the urban communities, affected significantly the urban dissatisfaction (Blorim and Hanke, 2005; Mishley et al., 2004; Liska et al., 1982; Skogan and Mazfield, 1981; Diazli; Benister and Fife, 2001, Shwartz et al., 1999; Pin, 2001:899) and many factors in the city have created and intensified it (Mishley et al., 2004). Urban space is structural in which human life is moving through it. So, human movement in this structure needs a space which is compatible with spiritual, psychological, and physical conditions of the residences. Urban space without its psychological security is just a sole communication artery; the presence of fear of crime in urban environment shows troubles of communities in modern age.

Fear is created in a person by understanding the potential danger of a place (whether explicitly or mentally) (Ferraro, 1995; Ferraro and Lagrang, 1987; cited by War, 2000) and it is a consequence of experiences, memories and relations with others (Koskela, 1997, cited by Yazoof, 2010). So, fear of crime includes a wide range of tentative and sensitive reactions to crime and disorders which individuals and communities are the creators of it (Hollway and Jefferson, 1997; Loader et al, 1998, cited by Pin, 2000).

Fear of crime in urban public places is a social issue. As public places form the most important part of the urban environment that include streets, squares, alleys, and all places where people have physical and visual access to (Tibalder, 1992; John Pirmore et al, 1994; Izzac Joseph cited by Fialkoof, 2004). Fear of crime is of important issues which reduces access to public places and restricts interaction with these places. Fear affects city form, urban and residential design and spatial distribution of the resident significantly (Lemanski, 2004:102; Benisterand Fife, 2001; Foster et al, 2010). In fact the relation between fear of crime and city and reinforcing this relation by some urban characteristics (population accumulation, racial and cultural inhomogeneity etc.) causes the resident of large cities to fear from facing crimes and this fear and anxiety restricts the individual's interaction with the space (Blobaum and Hunecke, 2005). In addition to investigation social factors, it has been tried in this research that spatial components affecting fear of crime in public places inside and outside the neighborhoods is considered. According to Harwy, any general theory about a city should be able to relate social processes of the city to its spatial form. The method of space formation can affect social processes deeply. In fact the spatial form of a place is a reflection of social relations. Based on his opinion, the role of space and location can be recognized in personal life by sociologic imagination or it can be found that how these relations between people and organizations are affected by the place which is separating them (Harwy, 1997). So, in order to build a bridge between sociological and geographical vision the relation between urban spatial form and the communicative behavior within it should be studied

Among urban public places, streets and their pathways are the most fundamental vital parts of the city. The entire city will be secure from fear and vandalism if the streets of the city are secured from fear and vandalism. Protecting city security is the main task of streets and pathways of a city (Jacobz, 2007) and the importance of the street as key element of forming a urban structure should be rebuilt (Tibaldz, 2006). Lack of fear and the sense of security in public places are necessary conditions of urban life. Tangible security of an environment is a necessary condition for attracting people to the sectors within the city (Ray Gindroz, cited by Erendet et al, 2008). According to Ellin (1997), public place is destroyed of people do not use a place out of fear or inconvenience.

On the other hand, one of the characteristics of poor and inefficient urban space is the capability of creating fear and anxiety. Following the presence of stress and pressure in the environment, perception range of the individual is reduced and vast part of logical thinking ability is lost and the learning ability is also decreased (Salingarous, 1999).

Fear of crime is one of the effective components of public health so that many studies are focused on its effect on welfare and personal well-being (Homlshim et al, 2010; Jackson and Stafford, 2009; Green et al, 2002). On the other hand, fear of crime in personal level reduces personal freedom (following the restriction of movement and activity), public communication, mutual trust and individual's social capital (Benister and Fife, 2001), causes anxiety and fear in the person, alienation and dissatisfaction in life and unnecessary protection and care of the people and in social level it reduces informal social control, transforms public streets of the city to dangerous places and decreases usage of urban areas, uniformity and coordination of the district and also reduces participation in neighborhood associations and social groups (Kaya and Kubat, 2007; Aram, 2009; Green et al, 2002; Kohm, 2009; Franklin et al, 2008; Black lee and Snider, 1997; Taylor, 1988 cited by Donges, 2000; Jackson and Gray, 2009; Grabosk, 1995; Wyant, 2008; Skogan and Maxfield, 1981; White, 1987 cited by Mac Cra et al, 2005; Mishley et al, 2004; Bannister and fyfe, 2001). Thus, fear of crime has the importance as the crime itself (Miller, 1973, cited by Louis and Salem, 1988) so that many people are affected by fear of crime rather than crime itself (Yaviz and Welch, 2010;; Schweitzer et al, 1999; Evanz and Fletcher, 2000).

The high feeling of insecurity among citizens requires more attention to the subject of the research. In Previous studies is less attention simultaneously to the action and structural factors affecting the fear of crime and according to the importance of social phenomenon of fear of crime and its harmful consequences and lack of special emphasis on spatial components on fear of crime, choosing this topic for the research seems necessary. The aim of this study is to answer the following questions: How much is the amount of fear of crime in public places of Mashhad city? What are the effective factors on fear of crime in public places of Mashhad city?

## *2. Empirical background*

Micsheli (2004), they research on fear of crime in Italy and the factors affecting it. Miceli (2004), He showed that fear of crime associated with the extent of the crime is broader than the crime itself. McCrea et al (2005) Explores the fear of crime and disorder and showed individual characteristics play an important role in explaining the fear of germ that are local, Schafer et al (2006), He showed that women more than men experience fear of crime. Kristjansson (2007), he did Comparative research on insecurity in the two European countries. fergosen and Mindle (2007) they study individual levels of fear of crime and factors affecting it. Franklin et al (2008), they studied models of vulnerability, disorder and social cohesion effective fear of crime model of irregularities was the best model to describe the fear of crime, to analyze and model of irregularities was the best model to describe the fear of crime, Kohm (2009) He study spatial data, victimization patterns and local experience to explain the fear of crime in Washington. Daglar (2009) with Using qualitative research and in-depth interviews among students explored the fear of crime, Nicholson (2010) with emphasis on the impact of family structure, to the conclusion that fear of crime had no significant effect on the family structure and local environment variables and are more effective. are some of the researchers who studied fear of crime, it is criteria, insecurity feeling and fear of victimization security condition in urban areas inside the country and measured and assessed personal and structural factors carefully.

## *3. Relevant literature linking fear of crime*

### *3.1 Theoretical framework*

In this section we first investigate convergence and divergence of theoretical and empirical records of fear of crime and then present our proposed definition of fear of crime and finally express effective factors on fear of crime and according to that we reveal our theoretical analysis model.

Environmental and special approach: This approach Based on this approach the key to understanding the fear of crime is how people experience and interpret urban areas and factors affecting fear of crime should be studied within three components of urban location (Canter, 1976)-as meaningful part of the city-, i.e. mental image of people from the place, people activity in the place and physical and compositional structure of the place. According to this approach people see the environment effectively as a measurement means for assessing danger and supportive factors. Environment provides people a possibility of visual confirmation of criminal danger probability. So fear is related to the city and the method of using urban areas and to its denotation (Banister and Fyfe, 2007). Newman believes that poor designing of urban areas and in fact the compositional structure of the urban area increases the opportunity for crime realization and also decreases the people territory and willing for using and defending district areas (cited by Pin, 2000). According to Newman (1972, 1973), residences of the district can have major role in decreasing the ground for crime committing in their district by using special methods. Many studies have approved that there is a relation between fear of crime and spatial view so that fear is more affected by spatial configuration rather than the crime itself (Kia and Koubat, 2007). The Queen Linch's theory (1960) of urban form meaning and Samoel Shamay's theory (1990) of location sense emphasize on the role of people's mental visions of the place for reducing fear of crime and insecure feeling. Linch (1960) emphasize on urban area legibility for providing security so that legible urban area let people communicate with it easily and recognize different parts of the space and they do not experience frustration and insecurity. According to the theory of "location sense" of Samoel Shamay (1991), dependency and attraction to a place reinforce the sense of emotional security in the human. So understanding and feelings of a person is tied to the meaning of the environment and is unified with it. Jackob's theory of street eyes (1961) emphasize on the role of people presence and movement in the area and also on variation of area functionality and visional penetration (windows and balconies view of the streets and public area) for providing security in urban areas.

According to the theory of broken window (1982), living among signs of disorder and indecency may lead to inducing anxiety, fear, anger, and depression. As the residences of these districts feel the threat and lack of interconnection among themselves. In this approach disorder means observable factors which represent violation of social order and control in the district. Tibaldz's theory of human-based city also outlines the role of compositional factors (visional penetration and creating various usages in public areas) and mental image of people from the place (legibility of the public place) to secure public areas.

### 3.2 Non-environmental approaches

According to this approach, factors other than environmental and spatial elements are affecting the formation of fear of crime. Among theories related to this approach, vulnerability theory, victimization theory and social control theory can be names. "Vulnerability theory" explains what provides the ground for fear in the people is inability and powerlessness against crime and defending from themselves. Sometimes this feeling is due to physical inability and sometimes due to standing in a certain social position. Economic and social positions of the people are important factors affecting social vulnerability. Those who live in delinquency areas or do not have suitable economic position to gain necessary supportive resources will have more insecure feelings and consequently experience more fear (Kilias and Klersi, 2000).According to the theory of "victimization", direct and indirect victimization (being informed about other's victimization by acquaintances and following crime news through media like visual, audio or written media) have significant effect on creating fear of crime (Bawmer, 1978; Taylor and Hall, 1986;

Gerbener et al, 1977, 1976). According to the theory of "social control", fear is determined in relation to the ability of the people to apply control on their living area and other's behavior and activities. Based on this approach people are afraid of something which is not able to prevent it or overcome victimization or when they feel inability. This approach creates a clear relation between urban life situations and incidence of fear of crime (Banister and Fyfe, 2007).

## 4. Factors affecting fear of crime

There are many factors affecting fear of crime. These variables can be investigated in two levels of agent and structure (minor and major) by structural theory of Gindez (Gindez, 1995). Personal and demographic variables are generally considered in the level of agent (minor) and social variables in the level of structure (major).

### 4.1 Reactive factor

One of the reactive variables affecting fear of crime is vulnerability. Some of the other affecting personal factors on the amount of fear of crime are gender and age. The effect of gender on the fear of crime is approved in many researches. In fact the fear of crime in women is more than men. However, victimization experience is more in men (Kristjansson, 2007; Mireku, 2002; Fergosen and Mindle, 2007; Tesloni and Zarafonito, 2008; Lee, 2001; Callanan et al, 2009; Myet al, 2010; Mishelly et al, 2004; Scaffer et al, 2006; Mac Craw et al, 2005; Ferraro, 1995; Daglar, 2009). Age is another effective variable among demographic variables. Although young people especially young men aged 16-24 years old are often at risk of crimes, older people report the highest level of fear of crime (Scogan, 1995; Mac Carel et al, 1997; Taylor and Hall 1986 cited by Schefer, 2006; Kristjansson). Mental assumptions of people related to their surroundings are of reactive factors affecting the amount of fear of crime. Location sense (Shamay, 1990) and location legibility (Linch, 1960; Tibaldz, 1992) have significant effect on reducing fear of crime as two types of implication understanding of the people about their surroundings. The amount of people presence and movement in public places (Jackobs, 1961; Tibaldz, 1992) which is an indication of people activity in the place is another reactive factor affecting the amount of fear of crime. Duration of residency in a place is also another reactive factor which affects theamount of fear of crime reductively (Mishley et al, 2004). Victimization also creates fear of crime which is in two forms of direct and indirect victimization (Bowmer, 1978; Taylor and Hall; Kristjansson, 2007; Kohm, 2009). Cumulative effectiveness of the neighborhood, including people participation in district social works and monitoring district al works, social informal control and trust and resident support of each other also affect reducing fear of crime (Loius, Salem, 1985; Hunter, 1986; Fergosen and Mindle, 2007; Kristjansson, 2007). According to Kent Ferraro theory, understanding the risk has significant role in creation of fear of crime (Ferraro, 1995; Kristjansson, 2007; Fergosen and Mindle, 2007).

### 4.2 Structural factors

Compositional and physical structure of a place including effective structural factors on fear of crime can be outlined as visual accessibility (Tibaldz, 1992; Noshahrgarayan, 1990) and disorder (Wilson and Kling, 1982) and non-defensive areas (Newman. 1972). The presence of various functions in the place is of characteristics of urban area compositional properties which increase people presence in the area and the security of the area is supplied by unconscious monitoring of the people (Jackobs, 1960; Tibaldz, 1992; Noshahrgarayan, 1990). Existence of undefended areas in the residential district (such as obsolete and abandoned lands and buildings, dark streets and recessed areas) which reduce visual view of a place and provide the ground for committing a crime also increase fear of crime (Newman, 1972). Disorder in public areas of the city is another spatial characteristic which affect fear of crime. Disorder in a society or a neighborhood increases vulnerability sense and anxiety stem from the crime. On the other hand, disorder increase will lead to the reduction of social coherence and also affect district satisfaction and cumulative effectiveness negatively (Fergosen and Mindle, 2007).

Location rank is another structural factor affecting the amount of fear of crime which means those who live in districts with higher economic-social position residents will have lower fear of crime (Kristjansson, 2007; Mishley et al, 2007).

## *5. Major hypothesis of fear of crime*

- Fear of crime is essentially affected by two classes of reactive and structural factors and the interaction between them. Now we study the relation between reactive also structural factors and Fear of crime.
- ➢ reactive factors
    - Vulnerability increases the fear of crime.
    - Victimization increases the fear of crime.
    - Understanding the risk of the crime increases the fear of crime.
    - People's mental images of a place decrease the fear of crime.
    - People's activities in a place decrease the fear of crime.
    - Cumulative effectiveness in a place decreases the fear of crime.
    - The more is the age ratio is in a neighborhood, the more will be the fear of crime.
    - The more is the gender ratio of women than men the more will be the fear of crime in the neighborhood.
    - The longer is the duration of residency the lower will be the fear of crime.
- ➢ Structural factors
    - Visual penetration of a place leads to the reduction of fear of crime.
    - Varying the usage of a place decreases the fear of crime.
    - Disorder in a place increases the fear of crime.
    - Undefended areas in a neighborhood increase the fear of crime.
    - Residential rank decreases the fear of crime.
- ➢ Interaction between structural and reactive factors
    - Victimization increases vulnerability sense.
    - Elders feel more vulnerability.
    - Women are more vulnerable than men and have more fear of crime.
    - Location disorder decreases cumulative effectiveness and increases the fear of crime.

Fear of crime in public places is considered as dependent variable in this research which includes fear of crime in public places inside and outside the neighborhood. Independent variable: includes two general categories of reactive and structural factors and the interaction between them as follows:

Reactive factors consist of vulnerability, victimization, gender ratio of the place, age ratio of the place, mental image of the place, activity in the place, cumulative effectiveness, and residency duration in the place and understanding the risk of the crime. Structural factors include visual penetration of the place, usage versatility of the place, undefended area, and site rank. The interaction between personal and structural factors is also considered as independent variable which includes: disorder and effectiveness, gender ratio and vulnerability, age ratio and vulnerability, site rank and vulnerability.

## *6. Research method*

The most important reason for select Mashhad in this research is Cultural diversity and ethnic heterogeneous. Mashhad has a disintegrate context. Mashhad is a religious metropolis with many job opportunities and suitable educational and health facilities like London. The difference is that Mashad evidently, attract many emigrants from internal Small towns and Villages also, since Mashhad have common boundary with Afghanistan, a large number of poor and jobless emigrant move to Mashhad. As a result, Mashhad become to full of crime metropolis. Consequently, foreign immigrants commit to different kind of crimes and make insecurity.

The five areas selected in this study were selected by these logical spectra that are expected to be a multicultural religious city. In other words, these elected areas contain firstly, Sajjad area with the capitalist class and wealthy stratum lack petty crimes. Secondly, Lashgar area with trades people, employees, and also foreigners mostly from Arabic countries of the Persian Gulf with relatively low level of petty crimes, Thirdly, koye Amir area with relatively poor and the marginalized stratum with relatively high level of petty crimes and eventually, Khajeh Rabi areas with poor people and temporary jobs had many petty crimes and Avini area with a large number of poor people and marginalized due to lack of jobs and the families of criminals, criminals and foreigner from war-torn countries such as Afghanistan and Iraq with very high level of petty crimes was studied. These areas respectively, zones 1, 10, 2, 3, and 5 of Mashhad Municipality also from each region selected randomly 10 neighborhoods and each neighborhood to collect data from 45 persons.

The method of the paper is surveying and the data collection tool is questionnaire. Sampling method is stratified non-proportionate. It means that by considering the sample (2250 household), 50 blocks (each block is assumed as a neighborhood) in 5 different rank clusters in Mashhad city and in each block 450 house hold were selected randomly. Blocks were also selected by stratified method and probability proportionate with the size method. In each block 400 households were selected orderly and a member of each family answered the questions who were above 18 years old and more. For all variables of the research, their factorial score is calculated by factorial analysis as weight criterion. Scale range is calculated 0-100 for all criteria. Dispersion statistics is used for descriptive purposes and one way ANOVA for comparing the average of research variables in different neighborhoods. Pearson's coefficient of contingency and general linear model-multivariate is used for data analysis.

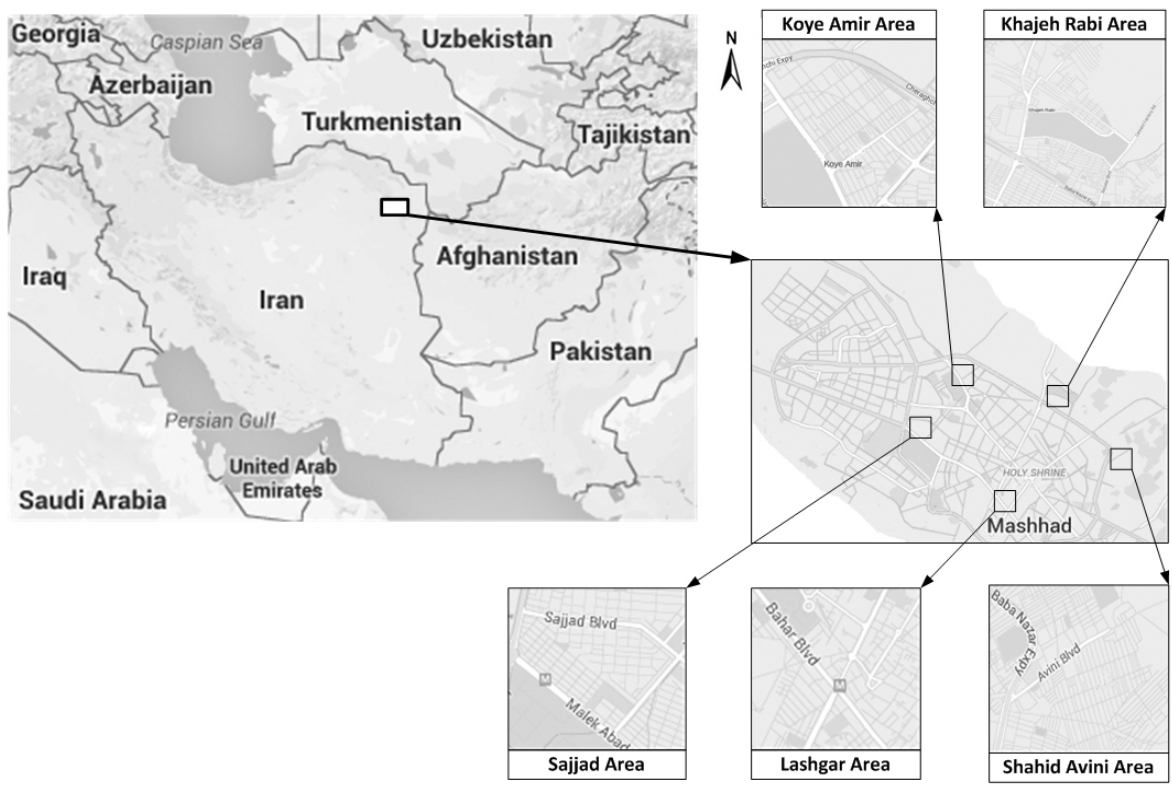

Figure 1: Five areas elected in Mashhad, Iran.

## 7. Results

In all neighborhoods men (43 percent) and women (57 percent) were included. Average of age in neighborhood was over 30 years old.

### *7.1Fear of crime description in public areas of Mashhad city*

The most common fear of crime in public places inside the neighborhoods was fear of housebreak (average 49.3) and the most common fear of crime in public places outside the neighborhood was fear of purse snatching (average 56.6). The lowest fear of crime in public places inside and outside the neighborhood is assault by others. As the average fear of assault inside and outside the neighborhood is 25.3 and 38.8 respectively. Fear of purse snatching (average 33.3) and fear of mugging and racketeering (average 31.3) are the most common fears in public places inside the neighborhood after housebreaking.

Below graphs show the amount of fear of crime in public places inside and outside the neighborhood, moreover, the figures indicate that which crime inside and outside areas have Higher frequency.

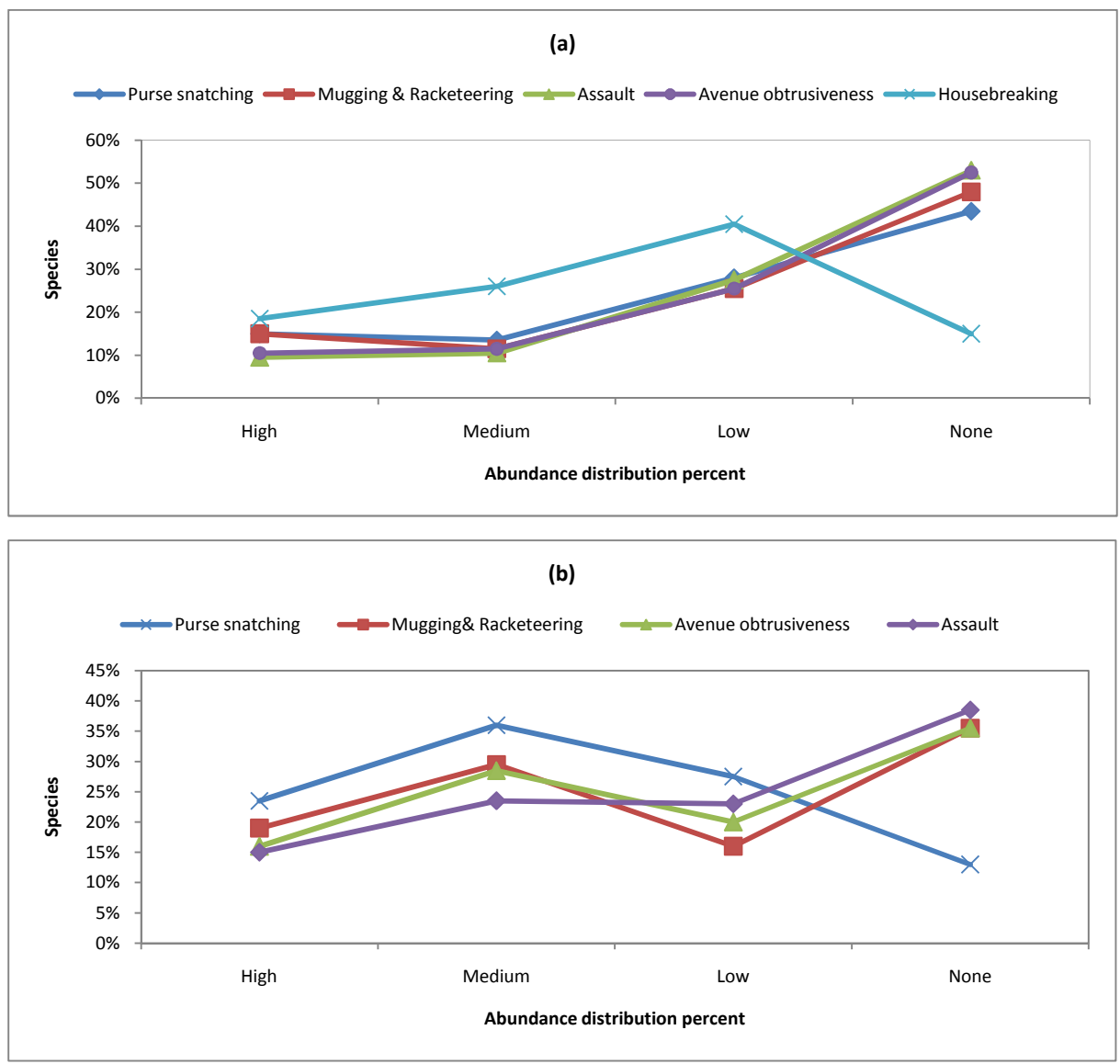

Figure 2: Fear of crime in public places (a) inside and (b) outside the neighborhood.

The comparison of average fear of crime in public places inside and outside the neighborhoods of different residential areas show that fear of crime in public places is varied for different neighborhoods.

Citizens afraid of pure snatching, mugging and racketeering, assault, avenue obtrusiveness in public places inside the neighborhoods more than public places outside the neighborhood. House breaking is a kind of crime that citizens experience just in public places inside the neighborhoods. All a result, citizens fell more security in inside their neighborhoods.

As has been noted in previous studies, house breaking variable is different another variables. Since when people walk either inside or outside the neighborhood or neighborhoods they lack defenses instrument and they are mostly alone while these days houses have been equipped with security systems, alarms and surveillance cameras, the ordinary citizens do not mobile these item with their selves.:

Fear of pure snatching in public places inside the neighborhood is more than public places outside the neighborhood. Because residents are more familiar relevant to their spatial and area residents and   if they have been attacked by a stranger more likely their neighbors help them moreover in public places inside the neighborhood local social capital is stronger ,but in public places outside the neighborhood people experience more fear due to anonymity of space and strangers residents and  less likely when they are at risk receive help from other.

Weighted mean of fear of crime in public places inside the neighborhoods of Lashgar and Khajeh Rabi district was more than other neighborhoods and was lower for very low neighborhood. Weighted mean of fear of crime in public places outside the neighborhoods of Lashgar and Khajeh Rabi district was also more than other neighborhoods but the lowest fear of crime in public places outside the neighborhoods belonged to the neighborhoods located in Sajjad district. On the other word, residents of Lashgar and Khajhe Rabi are more exposed to the fear of crime outside the neighborhood than others.

*Table 1: weighted mean of fear of crime in public places inside and outside the neighborhoods separated by residency district*

| Variable name | Residency | | | | | Significance level of difference of average fear of crime in different neighborhoods |
| --- | --- | --- | --- | --- | --- | --- |
| | Neighborhood of Sajjad district | Neighborhood of Lashgar district | Neighborhood of Koye Amir district | Neighborhood of Khajhe Rabi district | Neighborhood of Shahid Avini district | |
| Fear of crime in public places inside the neighborhoods | 26 | 44.6 | 32.8 | 41.6 | 20.6 | 0.000 |
| Fear of crime in public places outside the neighborhood | 31.2 | 58.9 | 45 | 45.4 | 45.2 | 0.000 |

Below graphs show which any offense is repeated within each region. In Sajjad area that it is wealthy and rich the residents experience fear of crime in public places inside the neighborhoods   is lower than in other areas. Residents of Khajeh Rabi area experience the highest level of Fear of crime expect fear of avenue obtrusiveness and Fear of assault. On the other hand, residents of Lashgar area with heterogeneous context allocated fear of avenue obtrusiveness and Fear of assault with the most frequent.

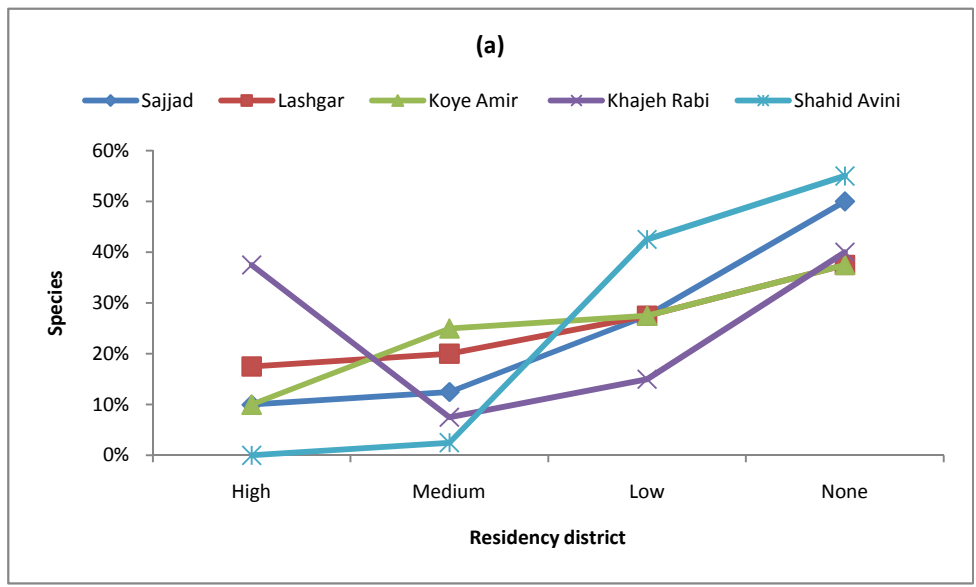

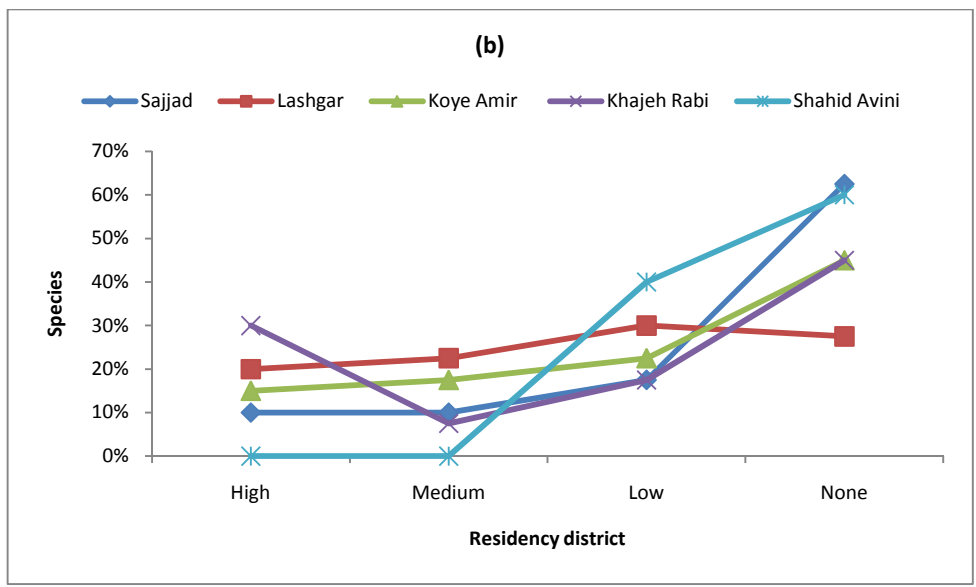

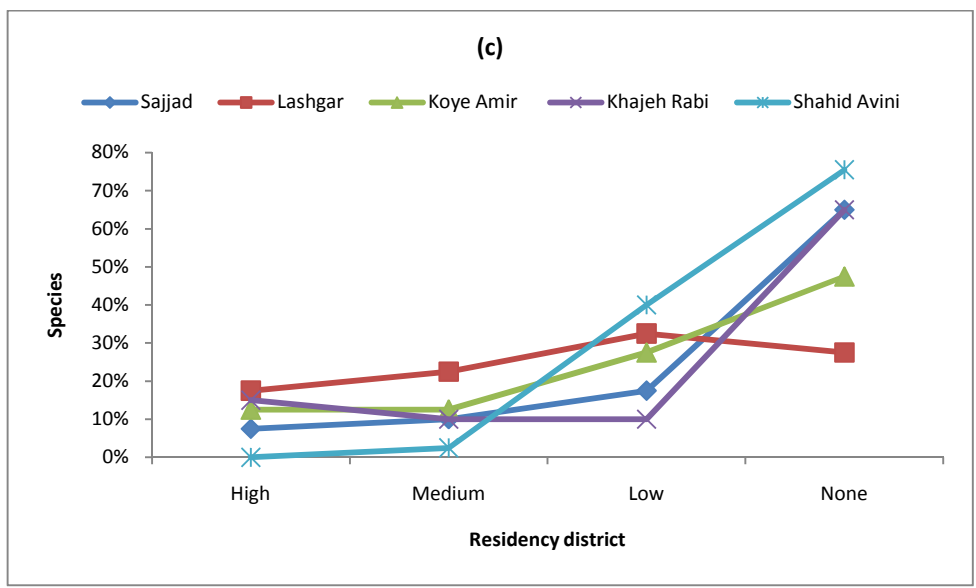

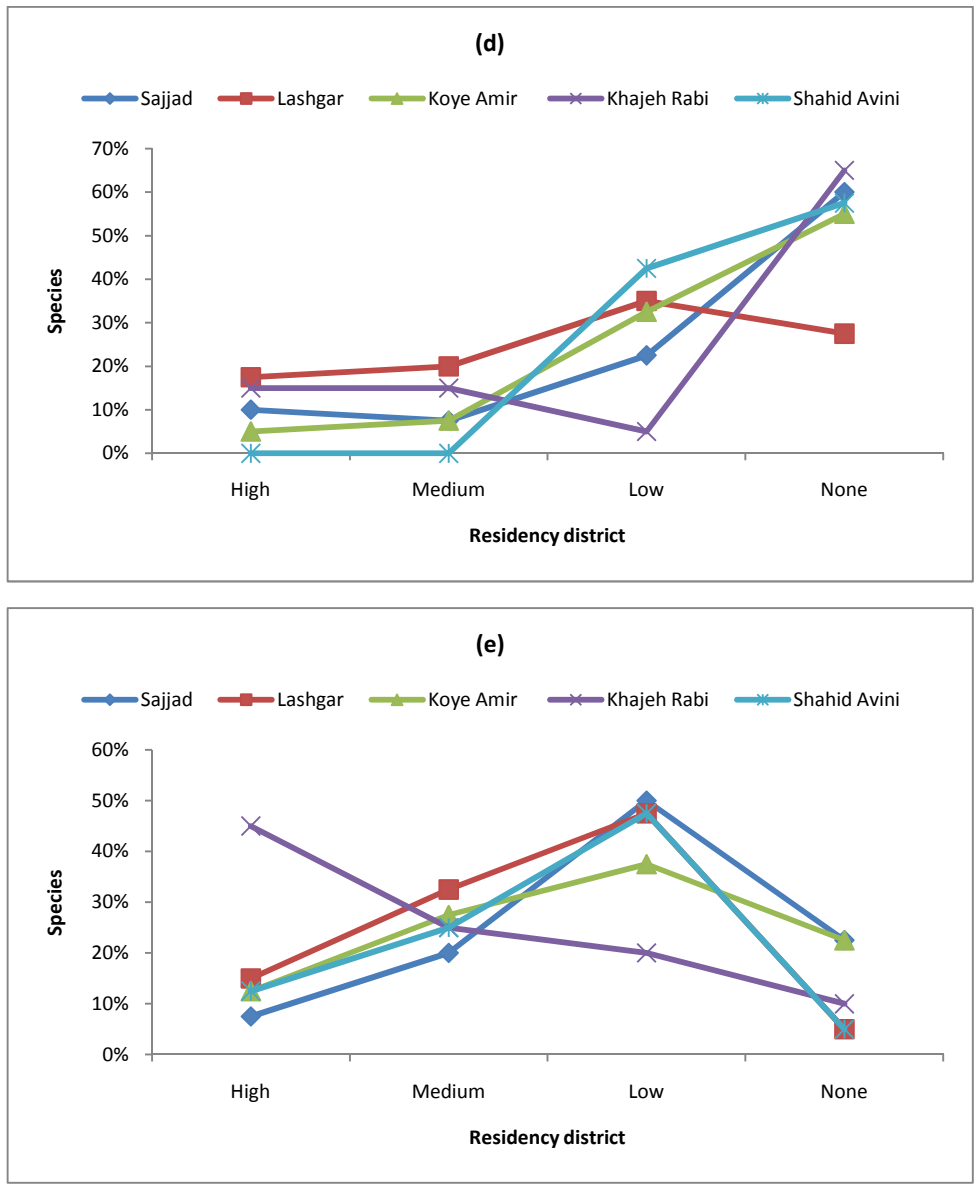

*Figure 3: Abundance percent of fear of crime in public places inside the neighborhoods separated by residency district, (a) Fear of purse snatching, (b) Fear of mugging and racketeering, (c) Fear of avenue obtrusiveness, (d) Fear of assault, (e) Fear of housebreaking.*

The graphs show all kind of crime are just over in Khajeh rabi area expect fear of avenue obtrusiveness and fear of assault. The most important reason, in favor of this matter can be that in lashgar area has different context. On the other word, resident in lashgar area are most pilgrim from another cities in Iran or tourist from Arabic country and since Arabs are often aggressive the rate of assault is high also Arabs like Iranian woman so they make avenue obtrusiveness as a consequence, people special Iranian woman feel insecurity. In Avini area nearly, we face fear of housebreaking and another crimes allocated low rate. Major reason can be that residents of there are often religious, calm,peaceful, honorable poor so they commit crime less than another area. Overall in Sajjad area that rich people leave there all crimes repeat less than another area on the other words, they commit petty crimes less than another area

### 7.2 Descriptive statistics of fear of crime in public places inside the neighborhood

Variable of fear of crime in public places inside the neighborhood has the mean of 33.1 with minimum and maximum scores of 0 and 100 respectively. Standard deviation of the variable of fear of crime in public places inside the neighborhood (16.3) shows that people's fear of crime in public places inside the neighborhood is very fluctuating. Half of respondents have a score lower than 34.1 and a quarter have a score more than 48.3 for the variable of fear of crime in public places inside the neighborhoods.

As the following table shows fear of crime in public places outside the neighborhood (45.1) is an indication of higher fear of crime of respondents in public places outside rather than inside the neighborhood. In other words, people outside of their own neighborhoods experience more fear of crime.

*Table 2: descriptive statistics of fear of crime in public places inside and outside the neighborhood*

| Variable name | Minimum | Maximum | Mean | Skewness | Standard deviation | Quartiles | | |
|---|---|---|---|---|---|---|---|---|
| | | | | | | First quartiles | Second quartiles | Third quartiles |
| Fear of crime in public places inside the neighborhood | 0 | 100 | 33.1 | 0.02 | 16.3 | 18.3 | 34.1 | 48.3 |
| Fear of crime in public places outside the neighborhood | 0 | 100 | 45.1 | 0.3 | 18.7 | 31.2 | 46.8 | 58.3 |

### 7.3 Describing structural factors affecting fear of crime

The following table shows descriptive statistics of spatial factors affecting fear of crime show. The results of the table are as follows: undefended areas of the studied neighborhoods are 46.9 percent. Visual accessibility of the investigated neighborhoods for the residents is lower than 50 percent and equals 46.6 percent. Usage versatility in studied areas is lower than 50 percent and the disorder observed in public places inside the neighborhoods is 49.8 present.

*Table 3: descriptive statistics of factors affecting fear of crime*

| Variable name | Minimum | Maximum | Mean | Skewness | Quartiles | | |
|---|---|---|---|---|---|---|---|
| | | | | | First quartiles (25) | Second quartiles (50) | Third quartiles (75) |
| Undefended areas | 0 | 100 | 31.5 | 0.003 | 22.2 | 31.9 | 38.8 |
| Visual penetration | 0 | 100 | 58 | 0.27 | 50 | 58.3 | 63.8 |
| Usage versatility in the area | 0 | 100 | 32.4 | 1.3 | 13.8 | 23 | 33.7 |
| Disorder | 0 | 100 | 35.7 | 0.13 | 28.3 | 36.6 | 41.6 |

Table 4 shows mean spatial characteristics of the studied neighborhoods. Spatial factors affecting fear of crime in Neighborhoods of Khajeh Rabi are more than other area. Results of mean comparison are as follows:

*Table4: mean of spatial factors affecting fear of crime separated by resident district*

| Variable name | Neighborhoods of Sajjad district | Neighborhoods of Lashgar district | Neighborhoods of Koy Amir district | Neighborhoods of Khajeh Rabi district | Neighborhoods of Shahid Avini district | Significance level of the difference of mean spatial factors in different neighborhoods |
|---|---|---|---|---|---|---|
| Undefended areas | 21.67 | 32.50 | 26.39 | 44.17 | 33.06 | 0.000 |
| Visual accessibility | 57.2 | 55.2 | 53.8 | 69.7 | 52.2 | 0.004 |
| Usage versatility | 53.3 | 13.8 | 23 | 55.8 | 16.1 | 0.000 |
| Disorder | 25.1 | 35.8 | 31.5 | 47.6 | 38.3 | 0.052 |

*Table 5: descriptive statistics of personal factors affecting fear of crime*

| Variable name | | Neighborhoods of Sajjad district | Neighborhoods of Lashgar district | Neighborhoods of KoyAmir district | Neighborhoods of Khajeh Rabi district | Neighborhoods of Shahid Avini district | Significance level of the difference of mean spatial factors in different neighborhoods |
|---|---|---|---|---|---|---|---|
| Vulnerability | | 55.2 | 37.5 | 38.6 | 55.5 | 75.5 | 0.000 |
| Understanding the risk of the crime | | 23.8 | 40.2 | 25.5 | 36.9 | 26 | 0.000 |
| Victimization | Direct experience of victimization | 0 | 0 | 0 | 0 | 0 | 0.000 |
| | Indirect experience of victimization | 1 | 1 | 0 | | 0 | 0.000 |
| Cumulative effectiveness in the place | Neighborhood cooperation | 14.7 | 12.2 | 16.8 | 18.7 | 7.2 | 0.000 |
| | Social capital of the neighborhood | 64.5 | 66.6 | 73.9 | 74.2 | 71.8 | 0.000 |
| | Informal social control | 56.3 | 59.4 | 58.6 | 75.5 | 49.2 | 0.000 |
| Mental image of the place | Legibility | 88.8 | 81.3 | 88.8 | 87.8 | 63.6 | 0.000 |
| | Sire sense | 51.6 | 55.4 | 66.4 | 61 | 26.4 | 0.000 |
| Activity in the place | Movement in the place | 51.6 | 55.4 | 66.4 | 61 | 26.4 | 0.000 |

## 7.4 Multi-variable analysis of fear of crime

In this section, multi-variant standard deviation is used to test the simultaneous and reciprocal effects of structural and personal (reactive) factors on fear of crime in public places inside and outside the neighborhood. Table 6 shows all variable have Significance level expect Legibility and Usage versatility

*Table 6: results of Play multi-variant test*

| Variable name | Significance level (sig) | Eta squared statistics |
|---|---|---|
| Site rate | 0.000 | 0.20 |
| Disorder | 0.01 | 0.05 |
| Presence in the area | 0.00 | 0.02 |
| Legibility | 0.11 | 0.02 |
| Usage versatility | 0.47 | 0.00 |
| Visual penetration | 0.00 | 0.14 |
| Site sense | 0.00 | 0.11 |
| Social capital of the neighborhood | 0.00 | 0.11 |
| Informal social control in the place | 0.03 | 0.04 |
| Neighborhood cooperation in the place | 0.00 | 0.13 |
| Undefended areas | 0.00 | 0.24 |
| Vulnerability | 0.01 | 0.09 |
| Understanding the risk of the crime | 0.00 | 0.35 |
| Gender ratio of the place | 0.06 | 0.03 |
| Direct victimization | 0.00 | 0.08 |
| Indirect victimization | 0.00 | 0.09 |
| Vulnerability * gender ratio of the place | 0.19 | 0.07 |
| Vulnerability * gender ratio of the place | 0.00 | 0.13 |
| Disorder * informal social control | 0.00 | 0.09 |
| Disorder * social capital of the place | 0.00 | 0.11 |
| Disorder * neighborhood cooperation | 0.00 | 0.09 |
| Residency duration | 0.00 | 0.09 |

*Table 7: results of the inter-categorical tests*

| Results of inter-categorical test (Fear of crime in public places (the neighborhood)) | | | | Significance level | Eta squared statistics |
|---|---|---|---|---|---|
| Final approved model | | | Inside | 0.00 | 0.90 |
| | | | Outside | 0.00 | 0.90 |
| Reactive factors | Vulnerability | | Inside | 0.00 | 0.08 |
| | | | Outside | 0.48 | 0.00 |
| | Direct victimization | | Inside | 0.00 | 0.08 |
| | | | Outside | 0.48 | 0.00 |
| | Indirect victimization | | Inside | 0.34 | 0.00 |
| | | | Outside | 0.00 | 0.05 |
| | Understanding the risk of the crime | | Inside | 0.03 | 0.02 |
| | | | Outside | 0.00 | 0.32 |
| | Image of places | Site sense | Inside | 0.00 | 0.02 |
| | | | Outside | 0.73 | 0.00 |
| | | Legibility of the place | Inside | 0.05 | 0.02 |
| | | | Outside | 0.61 | 0.00 |
| | | Neighborhood cooperation | Inside | 0.00 | 0.13 |
| | | | Outside | 0.00 | 0.83 |
| | Mass Effect in places | Informal social control | Inside | 0.12 | 0.01 |
| | | | Outside | 0.78 | 0.00 |
| | | Social capital of the neighborhood | Inside | 0.59 | 0.00 |
| | | | Outside | 0.00 | 0.10 |
| | | Activity in the place (presence in the place) | Inside | 0.00 | 0.11 |
| | | | Outside | 0.01 | 0.03 |
| | Age ration of the place | | Inside | 0.00 | 0.11 |
| | | | Outside | 0.94 | 0.00 |
| | Gender ration of the place | | Inside | 0.21 | 0.00 |
| | | | Outside | 0.00 | 0.2 |
| | Residency duration | | Inside | 0.00 | 0.09 |
| | | | Outside | 0.11 | 0.01 |
| | Visual penetration | | Inside | 0.79 | 0.00 |
| | | | Outside | 0.00 | 0.13 |
| | Undefended areas of the place | | Inside | 0.00 | 0.32 |
| | | | Outside | 0.00 | 0.28 |
| **Structural factors** | Disorder | | Inside | 0.04 | 0.02 |
| | | | Outside | 0.00 | 0.22 |
| | Site rank | | Inside | 0.00 | 0.32 |
| | | | Outside | 0.00 | 0.28 |
| **Interaction between reactive and structural factors** | Vulnerability * age ratio of the place | | Inside | 0.00 | 0.13 |
| | | | Outside | 0.89 | 0.00 |
| | Vulnerability * site rank | | Inside | 0.00 | 0.26 |
| | | | Outside | 0.00 | 0.28 |
| | Disorder * social capital of the place | | Inside | 0.14 | 0.01 |
| | | | Outside | 0.00 | 0.08 |
| | Disorder * informal social control | | Inside | 0.27 | 0.00 |
| | | | Outside | 0.00 | 0.07 |
| | Disorder * neighborhood cooperation | | Inside | 0.00 | 0.07 |
| | | | Outside | 0.30 | 0.00 |

*Table 8: parameters of reactive and structural factors affecting fear of crime*

| | | Variable name | | Significance level (sig) | Efficacy coefficient |
|---|---|---|---|---|---|
| **Variables affecting fear of crime in public places outside the neighborhoods** | **Reactive factors** | Vulnerability | | 0.00 | 3.9 |
| | | Indirect victimization | | 0.03 | 2.1 |
| | | Understanding the risk of the crime | | 0.00 | 5.06 |
| | | Mental image of the place | Site sense | 0.00 | -5.02 |
| | | Cumulative effectiveness | Neighborhood cooperation | 0.00 | -5.02 |
| | | Activity and presence in the place | | 0.00 | -5.6 |
| | | Age ration of the place | | 0.00 | 4.7 |
| | | Residency duration | | 0.00 | -4.2 |
| | **Structural factors** | Undefended areas | | 0.00 | -5.6 |
| | | Disorder | | 0.04 | 2 |
| | | Site rank (residency in Khajeh Rabi district) | | 0.00 | -5.6 |
| | | Site rank (residency in Shahid Avini district) | | 0.01 | -2.5 |
| | **Interaction between structural and reactive factors** | Vulnerability * age ration of the place | | 0.00 | 0.04 |
| | | Vulnerability * site rank (residency in Sajjad neighborhood) | | 0.00 | -2.2 |
| | | Vulnerability * site rank (residency in Shahid Avini neighborhood | | 0.00 | -3.5 |
| Variables affecting fear of crime in public places inside the neighborhoods | Reactive factors | | Disorder * Neighborhood cooperation | 0.00 | -3.7 |
| | | | Direct victimization | 0.00 | 3.9 |
| | | | Indirect victimization | 0.00 | 3.1 |
| | | | Understanding the risk of the crime | 0.00 | 8.9 |
| | | Cumulative effectiveness | Social capital of the neighborhood | 0.00 | -4.4 |
| | | | Activity and presence in the place | 0.01 | -2.4 |
| | Structural factors | | Undefended areas | 0.00 | 4.1 |
| | | | Visual penetration of the place | 0.00 | -5.2 |
| | | | Site rank (residency in Khajh Rabi district) | 0.00 | 3.1 |
| | | | Site rank (residency in Shahid Avini district) | 0.04 | 2.07 |
| | Interaction of reactive and structural factors | | Vulnerability * site rank (residency in Khajeh Rabi district) | 0.00 | -2.9 |
| | | | Vulnerability * site rank (residency in Shahid Avini district) | 0.04 | -2 |
| | | | Disorder * informal social control | 0.00 | -3.7 |
| | | | Disorder * social capital of the neighborhood | 0.00 | -4 |

Variables not incorporated into the models of fear of crime in public places inside and outside the neighborhoods are as follows: there has been no significant relation between variables of direct victimization, site eligibility, social capital of the neighborhood, informal social control and gender ratio of the place from reactive factors and variables of usage versatility, visual penetration of the place fro, structural factors and fear of crime in public places inside and outside the neighborhoods in the model of fear of crime. Also the interaction between vulnerability and gender ratio, disorder and informal social control and disorder and social capital of the neighborhood has no effect on fear of crime in public places inside the neighborhoods.

In fear of crime model, variables of vulnerability, site legibility, site sense, informal social control, neighborhood cooperation, age ratio of the place and residency duration from reactive factors and variables of usage versatility from structural factors had no significant relation with the fear of crime in public places inside the neighborhoods. Also the interaction between variables of vulnerability and gender and age ratio of the place, disorder, and neighborhood cooperation has no effect on fear of crime in public places inside the neighborhoods.

## 8. Discussion

Undefended areas in the place (as one of the compositional characteristics of the place), site sense (one of the people's mental images from the place) and activity and the presence of the people in the place are three basic components forming the concept of a place and have significant effect on fear of crime inside the neighborhoods. Findings showed that the more people understand the risk of crime incidence in their neighborhood, the more their fear of crime in public places outside the neighborhood will be. In multi-variant analysis, victimization experience is also predictor of fear of crime in public places inside and outside the neighborhood Social-economic position of a person is a factor affecting fear of crime. This variable is a reverse predictor of their fear of crime in public places inside or outside the neighborhood. Residency area had the most prediction ability concerning fear of crime in public places inside and outside the neighborhood. So, in public place having higher level of informal social control fear of crime were lower inside and outside the neighborhood.

Understanding and interfering of the neighbors from informal social control capacity is a preventive measure related to fear of crime and insecurity feeling. Those who live in neighborhoods with higher fear of crime do not consider themselves competent for controlling criminal behavior of others. Social capital of the neighborhood is proved as another structural factor affecting fear of crime in public places inside the neighborhood. This means in public places where residents have more social capital (trust and support) towards each other fear of crime is lower inside the neighborhood. The effect of spatial factors on fear of crime in public places inside and outside the neighborhood shows that compositional and spatial factors of the living area affect fear of crime significantly. So, there is a relation between fear of crime and spatial view and fear of crime is affected by spatial configuration.

Visual accessibility and lighting of public places also cause fear of crime to decrease for the people inside the neighborhoods. The relation between movement and the presence of people in the place and fear of crime in public places inside the neighborhoods is proved. It means fear of crime becomes lower by increasing usage versatility and consequent people presence in public places inside the neighborhoods. Among studied structural and reactive factors, site rank has the greatest effect on the fear of crime in public places inside the neighborhoods. It means that fear of crime is decreased by increasing site rank of people's residency. And the effect of residency on fear of crime is approved in previous researches (2008; Kristjansson, 2007; Mishely et al, 2004). Findings showed that the more people understand the risk of crime incidence in their neighborhood, the more their fear of crime in public places outside the neighborhood will be. Also this is supported by previous researches (Kristjansson, 2007; Fergosen and Mindle, 2007) and according to Ferraro (1995) understanding the risk of the crime is one of the main determining factors of crime rather than the crime itself. In multi-variant analysis, victimization experience is also predictor of fear of crime in public places inside and outside the neighborhood which previous researchers (Duglar, 2009; Kohm, 2009; Kristjansson, 2007; Fergosen and Mindl, 2007) also supported it. Although there is a difference that direct victimization experience predicts more fear of crime in public places inside the neighborhoods and indirect victimization experience predicts more fear of crime in public places outside the neighborhoods. According to Bawmer (1978) and Scogan (1978), fear of crime is created in a person by personal experience of victimization or being informed of other's victimization.

Social-economic position of a person is a factor affecting fear of crime which is proved in previous researches (Schefer et al, 2006; Mishely et al, 2004). Social-economic position of a person is a reverse predictor of their fear of crime in public places inside or outside the neighborhood. As Scogan and Max Field stated (1981), those who feel more vulnerability than others, feel more fear of crime and insecurity. It should be noted that reciprocal effects of people's social-economic position and direct and indirect victimization on fear of crime in public places inside the neighborhood in multi-variant analysis were significant. It means that in studied areas where people had lower social-economic position and direct experience of victimization fear of crime were higher. As this has been proved in previous researches studies of (Kristjansson, 2007; Fergosem and Mindle, 2007, Louis and Salem 1986 and Taylor 1997 ) confirmed results of this research that Residency area had the most prediction ability concerning fear of crime in public places inside and outside the neighborhood. So, in public place having higher level of informal social control fear of crime were lower inside and outside the neighborhood. Social capital of the neighborhood is proved as another structural factor affecting fear of crime in public places inside the neighborhood. This is proved in theoretical and empirical backgrounds (Fergosen and Mindle, 2007; Agno, 1985). Variable of site legibility affected fear of crime in public places inside the neighborhood reductively. According to Queen Linch (2002, 2005), acceptable image of the place and legible view of the place for the person create secure feeling and avoid confusion. Visual accessibility and lighting of public places also cause fear of crime to decrease for the people inside the neighborhoods. As according to Tibaldz (2002, 2006) and Car (1992), lighting and visual accessibility of public places increase spatial clarification for the residents and following that they decrease fear of crime.

The relation between movement and the presence of people in the place and fear of crime in public places inside the neighborhoods is proved. As Francis Tibalds (2002, 2006) and Jean Jacobs also emphasized the key role of usage versatility of the place and increasing the presence of people in the place to reduce fear of crime and insecure feeling. According to them, the presence of people in the place is like assigning invisible eyes to control and manage the place. Among the studied spatial factors, only the relation of resident permanency and area identity with fear of crime outside the neighborhoods is proves. In other words, resident permanency and identity stability of public places lead to reduction of fear of crime in public places outside the neighborhoods. This means fear of crime of resident in places with the records of changing name and identity of the streets and alleys and multiplicity of home moving is more in places outside their neighborhood. The effect of residency duration is also proved in previous researches (Mishely et al, 2004).

## *9. Conclusion*

Fear of crime is an especially a problem that has troubled the urban communities, affected the urban dissatisfaction and many factors in the city have created and intensified it fear of crime has a destructive impact on social capital. Fear of crime lead to strengthen the sense of insecurity in society. Fear of crime is of important issues which reduces access to public places and restricts interaction with these places; reduce social trust and social participation. The research method is surveying and the information collection technique is through questionnaire. Probable span sampling (PPS) method is used. Sample population was 2000 households which were selected randomly in five categorical clusters from Mashhad city. The most important reason for select Mashhad in this research is Cultural diversity, disintegrate context and ethnic heterogeneous. We try to select 5area according with regarding wide range of citizens from rich and poor.

In this research we try In addition to investigation social factors, it has been tried in the research that spatial components affecting fear of crime in public places inside and outside the neighborhoods is considered. As a consequence, effective variable that have power of explain the fear of crime include victimization experience ,Social-economic position of a person, informal social control, Visual accessibility and lighting of public places, movement and the presence of people in the public place and t he stability of residence and identity space. According to recognize affecting factors on fear of crime due to reduction that and prevention of adverse effects it is better to take action in this regard done such as Increasing usage versatility of the place such as residential, business and recreational usage to enhance the presence of people in public places. Improving and enhancing visual accessibility of the pathways in public places such as lighting of the pathways, reinforce the neighborhood identity and site sense of the resident, securing public spaces and holding training classes for vulnerable groups such as women, elders and teaching feeling management and correct decision making in the context of facing the crime and the criminal.

## *Appendix*

In this part due to better understanding and clarity we show applying variables and scaling also Validity and reliability of the scale.

***Appendix A: applying variables and scaling.***

Table A1: (part 1).

| Variable type | Dimension | Elements | Empirical indicators |
|---|---|---|---|
| Locational component of the neighborhood (independent) | Activity in the place | Presence of people in the place | Walking in the neighborhood |
| | | | Daily shopping in the neighborhood |
| | Physical and compositional structure of the place | Visual penetration | Lighting of the pathways |
| | | | View of the home's windows to the streets |
| | | | The amount of seeing the beginning and end of the streets |
| | | Usage versatility of the place | Number of recreational, educational, business, health care centers, offices and green spaces in the neighborhood |
| | | Undefended area | Number of obsolete homes and lands, recesses areas and dead ended streets |
| | | Disorder | Presence of straying people, pile of garbage in pathways, noisy neighbors, abandoned cars, obsolete lands |
| | | Place legibility | Number of alley's, main and sideway street's signs, the possibility of connection between adjacent alleys, number of home plaque, |
| | | Site sense | Interest in continuing residency in the place, no willing for moving out of the place, comfortable sense of being in the place by remembering positive memories of living in that place, attempt to solve the problems of the place and devoting personal comfort for the |

| | | | neighborhood comfort |
|---|---|---|---|
| **Understanding the risk of the crime (independent variable)** | ________ | ________ | Probability of facing purse snatcher, mugging, racketeering, avenue obtrusiveness, avenue assault |

*Table A2: (part 2).*

| Variable type | Dimension | Elements | Empirical indicators |
|---|---|---|---|
| **Fear of crime (independent variable)** | Fear of crime in public places inside the neighborhood | ________ | Fear of housebreaking, purse snatching, mugging, racketeering, avenue obtrusiveness, avenue assault |
| | Fear of crime in public places outside the neighborhood | ________ | Fear of purse snatching, mugging, racketeering, avenue obtrusiveness, avenue assault |
| **Victimization experience** | Direct experience of victimization | ________ | Facing or not facing with direct crimes such as robbery, purse snatching, mugging, racketeering |
| | Indirect experience of victimization | ________ | Being informed of victimization of friends, acquaintances and relatives. |
| **Vulnerability (independent variable)** | ________ | ________ | Trust rate of neighborhood people, amount of recognition of residents from each other |
| **Cumulative effectiveness in the place (independent variable)** | Social control | Informal social control | Watching the stranger's communication, amount of recognition of neighborhood residents |
| | Social capital of the neighborhood | Trust | Trust rate of neighborhood people, amount of recognition of residents from each other |
| | | Support | Assistance of the resident during hard conditions |
| | District participation | ________ | Membership in cultural associations, Basij, masque of the neighborhood, sport team and cultural center of the neighborhood and participation in solving the neighborhood's problems |
| **Residency duration in the place (independent variable)** | ________ | ________ | Residency duration of people in the neighborhood |
| **Site rank (independent variable)** | ________ | ________ | Residency district of the person |

### Appendix B: Validity and reliability of the scale

Nominal and constructional credit was used to provide reliability. To do this factorial loads which are arranged based on exploration method for dependent variable are presented in below table. Pre-test is done by using 50 questionnaires to provide justifiability so that precision and accuracy of the measurement tools are ensured. The report of justifiability of main scales is presented in below table.

*Table B1: Factorial analysis of dependent and independent variables*

| Variable name | Variable type | | Sample number | Kayzer criterion | Bartlet statistics | Factor numbers |
|---|---|---|---|---|---|---|
| Fear of crime | Dependent | | 9 | 0.808 | 0.000 | 2 |
| Fear of crime inside the neighborhood | Dependent | | 5 | 0.847 | 0.000 | 1 |
| Fear of crime outside the neighborhood | Dependent | | 4 | 0.806 | 0.000 | 1 |
| Compositional structure of the place | Disorder | Independent | 3 | 0.628 | 0.000 | 1 |
| | Undefended area | Independent | 3 | 0.628 | 0.000 | 1 |
| | Visual penetration | Independent | 3 | 0.628 | 0.000 | 1 |
| | Activity in the place | Independent | 3 | 0.541 | 0.000 | 1 |
| Activity in the place | ——— | Independent | 2 | 0.500 | 0.05 | 1 |
| Mental image of the place | Legibility | Independent | 3 | 0.602 | 0.000 | 1 |
| | Site sense | Independent | 6 | 0.790 | 0.000 | 1 |
| Understanding the risk of a crime | ——— | Independent | 8 | 0.797 | 0.000 | 1 |
| Vulnerability | ——— | Independent | 3 | 0.659 | 0.000 | 1 |
| Cumulative effectiveness | Social control | Independent | 3 | 0.659 | 0.000 | 1 |
| | Social capital of the neighborhood | Independent | 9 | 0.823 | 0.000 | 2 |
| | Neighborhood participation | Independent | 6 | 0.865 | 0.000 | 1 |

*Table B2: justifiability coefficients of the scale*

| Row | Scale name | Element number | Justifiability coefficient | Row | Scale name | Element number | Justifiability |
|---|---|---|---|---|---|---|---|
| 1 | Fear of crime inside the neighborhood | 5 | 0.901 | 10 | Neighborhood participation | 6 | 0.835 |
| 2 | Fear of crime outside the neighborhood | 4 | 0.914 | 11 | Visual penetration of the place | 3 | 0.720 |
| 3 | Informal social control | 3 | 0.804 | 12 | Disorder | 6 | 0.748 |
| 4 | Vulnerability | 3 | 0.826 | 13 | Movement in the area | 2 | 0.753 |
| 5 | Site sense | 6 | 0.803 | 14 | Undefended area | 3 | 0.744 |
| 4 | Understanding the risk of the crime | 8 | 0.855 | 15 | Site legibility | 3 | 0.624 |
| 7 | Social capital of the place | 3 | 0.756 | | ——————— | | |

# References

-Daglar Murat. (2009)." Acomparative Study of fear of Sexual assult and personal property theft between international and noninternational students on an urban university campus".A dissertation for the degree of Doctor of Education in Leadership Education. Louisville, Spalding University,College of Education.

-Erendetm Rendel et al, (2008), principles of modern urbanism, translated by Aireza Danesh and Reza Basiry Mozhdehi, Tehran: publication of civil process and planning.

- Fialkoof, Yankel, (2004), urban sociology, translated by Abdolhossein Nik Gohar, Tehran: publication of Agah.

-Ferguson, Kristin M. Mindel ,Charles H.( 2007)"Capital Theory Modeling Fear of Crime in Dallas -Neighborhoods : A Test of Social Capital Theory", Crime & Delinquency, vol 53: 322.

- Ferraro,Kenneth.F(1995).Fear of Crime interpreting victimization Risk, New York: SUNY Press.

- Franklin Travis W.;Cortney A. Franklin and Noelle E. Fearn .(2008). A Multilevel Analysis of the Vulnerability, Disorder, and Social Integration Models of Fear of Crime. Soc Just Res; 21:204–227.

- Green, Geoff. Jan. M. Gilberston. Michael. J. Grimsely. (2002).Fear of crime and health in residential tower blocks: A case study in Liverpool,UK.European Journal of public health;12:10-15.

- Gindez, Anthony, (1999), modernism and status; society and personal identity in new age, translated by Mohsen Salasi, Tehran: Ney publication.

- Gindz, Anthony, (1998), modernism consequences, translated by Mohsen Salasi, Tehran: center publication.

-Jacobs, Jean, (2007), death and life of large cities of America, translated by Hamidreza Parsi, Arezoo Aflatooni, Tehran: Tehran University.

- Kohm, Steven .A.(2009). "Spatial Dimensions of Fear in a High-Crime Community: Fear of Crime or Fear of Disorder "Canadian Journal of Criminology and Criminal Justice.

-Kiwi, Rimon; Compenhood, Look Van, (1997), research method in social science, translated by Abdolhossein Nik Gohar, Tehran: Tootia.

- Leng, John, (2007), urban designing: typology of the approaches and plans along with more than 50 especial cases, translated by Hossein Bahraini, Tehran: Tehran University.

-Lemanski .Charlotte. (2004). A new apartheid? The spatial implications of fear of crime in Cape Town, South Africa. *Environment and Urbanization* 16: 101.

- Linch, Queen, (2002), urban form theory, translated by Hossein Bahraini, Tehran: Tehran University

- Linch, Queen, (1995), city face, translated by Manoochehr Mazini, Tehran: Tehran University Liska, A. L., Lawrence, J. J., & Sanchirico, A. (1982). Fear of crime as a social factSocial Forces,Vol. 60, No. 3pp. 760-770

- Lindstrom, Martin;Juan Merlo and Per-Olof.Ostergren.(2003) . Social capital and sense of insecurity inthe neighbourhood: a population-based multilevel analysis in Malm . o, Sweden. Social Science & Medicine 56:1111–1120.

- Madani Poorm Ali, (2000), urban area designing: principles of social and locational processes, translated by Farhad Mortezaee, Tehran: urban process and planning firm.

- May ,David C. Rader Nicole E. and Goodrum, Sarah.(2010)."A Gendered Assessment of the ''Threat of Victimization'': Examining Gender Differences in Fear of Crime" Perceived Risk, Avoidance, and Defensive Behaviors" .Criminal Justice Review.vol35: 159.

-McCrea Rod, Shyy Tung-Kai, Western John and Stimson .J.Robert. (2005). "Fear of crime in Brisbane; Individual, social and neighbourhood factors in perspective". Journal of Sociology. vol 41pp7-27.

Miceli,Renato,Roccato,Michele,Rosato,Rosal.(2004)."fear of crime in Italy :Spread and Determinants".Environment and Behavior. vol36:776.

-Nicholson,David F.(2010). "Disadvantaged neighborhood and fear of crime:does family structure matter?" A dissertation for the Degree of Doctor of Philosophy,University of Oklahoma,Graduate College.

- Pirmore, John et al, (1994), urban spaces, translated by Hossein Rezaee, Tehran: Tehran municipality CH.

-Pain. Rachel. (2001). Gender, Race, Age and Fear in the City. Urban Studies, Vol. 38, Nos 5–6, 899–913.

- Pain,Rachel ,(2000), Place, social relationsandthe fear of crime: a review, Progress in Human Geography vol24,3 pp. 365–387.

- Park, Andrew Jaehyung.(2008).Modeling the Role of Fear of Crime in Pedestrian Navigation . A, Dissertation for Degree of Doctor of Philosophy. School of Interactive Arts and Technology. Simon Fraser University.

- Ritzer, George, (1995), the theory of sociology in contemporary era, translated by Mohsen Salasi, Tehran: scientific.

-Schafer A. Joseph, Huebner.M. , Beth, Bynum. S. Timothy,(2006)."Fear of crime and criminal victimization: Genderbased contrasts",Journal of Criminal Justice ,34:285 – 301.

- Tankis, Feren, (2009), space, city and social theory, translated by Hamidreza Parsi, Arezo Aflatooni, Tehran: Tehran University Publication.

-Tibaldz, Francis, (2002), citizen-oriented urbanism, translated by Mohamma Ahamdinezhadm Esfahan: Khak.

- Town, Yee. Foo, contrast of the place sense and originality, (2005), human-based cities: improving public areas in large and small cities, translated by Hosseina Ali Leghaee, Firooze, Jadli, Tehran: Tehran University.